\documentstyle[12pt,epsfig]{article}
\def\jpsi{\mbox{J/$\psi~$}}
\def\pom{{I\!\!P}}            %gives pomeron symbol
\begin{document}
\begin{titlepage}
\hspace*{\fill}{IMSc-97/05/16}
\vspace*{\fill}
\begin{center}
{\Large \bf HERA Physics - an overview of experimental and
theoretical results}
\\[1cm]
Rahul Basu
\footnote{Presented at the XII DAE Symposium on High Energy
Physics, 26 Dec 1996 -- 1 Jan 1997 at Guwahati, India}\\
{\em The Institute of Mathematical Sciences, Chennai (Madras) 600 113, India}
\end{center}
\vspace{2cm}
\begin{abstract}
In this talk I review the low x QCD experimental results from HERA  
and their theoretical underpinnings. In particular, I discuss the structure
function $F_2$, large rapidity gap events, pomeron structure
functions and $J/\psi$ production. 
\end{abstract}
\vspace*{\fill}
\end{titlepage}
\section{Introduction}
The electron (positron)-proton collider facility at DESY, nicknamed HERA (for
Hadron Electron Ring Accelerator) has, since its inception in 1992, given
many new and interesting results from the two experiments H1 and ZEUS. These
experiments have thrown up a wealth of new theoretical issues pertaining to low
$x$ QCD. These experiments cover both the deep-inelastic region (DIS) and the
photoproduction region.
\begin{figure}[htb]
\begin{center}
\mbox{\epsfig{file=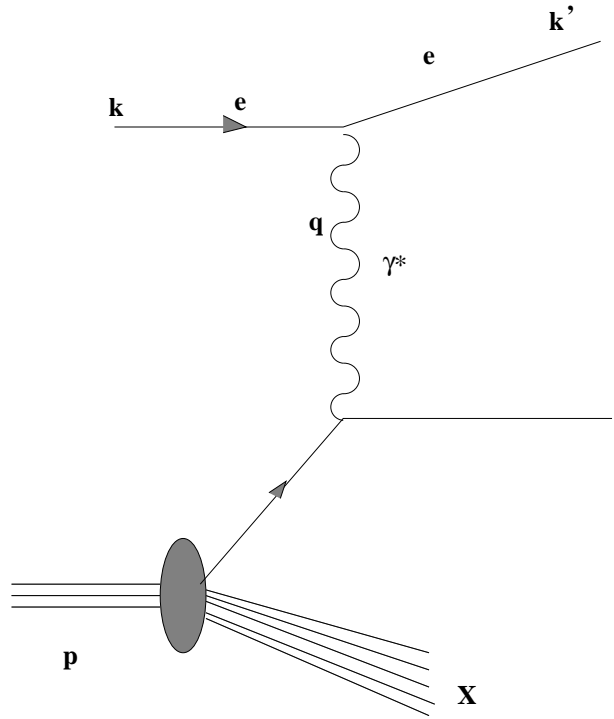,width=8truecm,angle=0}}
\caption{The basic DIS diagram with kinematic variables labelled}
\label{Fig 1.}
\end{center}
\end{figure}
The basic interaction picture is the usual DIS diagram shown in figure 1 for
the process
$$
e(k) + p(p) \rightarrow e(k^\prime) + X.
$$

The following kinematic variables completely describe the process:
\begin{eqnarray}
s=(k+p)^2 & \simeq & 4E_eE_p \\
Q^2 = -q^2 & \simeq & 2E_eE_{e^\prime}(1+cos\theta) \\
y=\frac{p.q}{p.k} & \simeq & 1 - \frac{E_{e^\prime}}{2E_e}(1-cos\theta) \\
x=\frac{Q^2}{2p.q}& \simeq & \frac{Q^2}{ys} \\
W=(q+p)^2 & \simeq & -Q^2 + ys 
\end{eqnarray}
Here $s$ is the CM energy squared, $E_e(E_e^\prime)$ is the energy of the
initial (scattered) electron, $\theta$ the angle of the scattered electron in
the lab frame with respect to the proton direction, $Q^2$ is the negative of
the momentum transfer squared to the proton, $y$ is the fraction of the
electron energy carried by the virtual photon in the proton rest frame, the
Bjorken variable $x$ is the fraction of the proton energy carried by the
struck quark and $W$ is the hadronic invariant mass. The masses of all
particles are neglected.

At HERA, a 27 GeV electron beam collides head-on with an 820 GeV
proton beam giving 
$$
s \simeq 4E_eE_p \sim 10^5 GeV^2
$$
which is much larger than hitherto obtained at fixed target
experiments. (Since 1994, the electron beam has been replaced by a
positron beam at the same energy). As a result the two experiments H1 
and ZEUS can measure the
structure function $F_2$ in a completely new $(x,Q^2)$ domain (see
Figure 2).
\begin{figure}[htb]
\begin{center}
\mbox{\epsfig{file=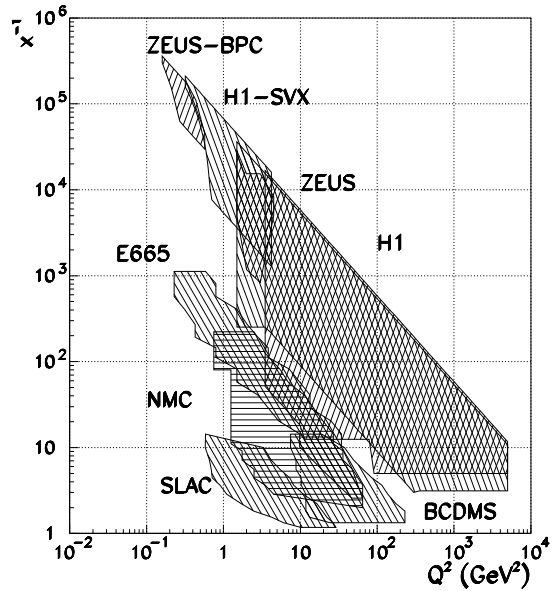,width=8truecm,angle=0}}
\caption{Phase space coverage of the $F_2$ measurements
(from \protect \cite{halina})} 
\label{Fig 2.}
\end{center}
\end{figure}

The Born cross section for single photon exchange in DIS is given by
\begin{equation} 
\frac{d^2\sigma}{dxdQ^2}=\frac{2\pi\alpha^2}{Q^2x}\left[2(1-y)+\frac{y^2}
{1+R}\right]F_2(x,Q^2)
\end{equation} 
where $R$ is the photoproduction cross section ratio for longitudinal
and transverse polarised photon, $R=\sigma_L/\sigma_T$. $R$ has not
been measured at HERA; therefore a QCD prescription based on
parametrisation of parton densities is used. $Q^2$ is directly
measured from the scattered electron but $x$ is calculated from $Q^2$
and $y$ and therefore depends on the experimental resolution of $y$.

\section{Measurement of $F_2^p(x,Q^2)$}
The data for $F_2^p$ from H1 and ZEUS \cite{h1,zeus} is shown in 
figures 3, 4, and 5, along with some of the older fixed target data. It is clear
that at fixed $Q^2$, $F_2$ rises with decreasing $x$ down to the
smallest $Q^2$, the steepness of the rise decreasing with decreasing
$Q^2$. Similarly at fixed $x < 0.1$ $F_2$ rises with $Q^2$, the rise
becoming steeper as $x$ decreases. (We shall henceforth drop the superscript $p$ on
$F_2$ since all our discussion henceforth will refer to the structure function of the
proton). In fact, as seen in figure 4, upto
$Q^2 \sim 0.85 GeV^2$ the data favour a parametrisation based on a
soft pomeron  such as that suggested by Donnachie and Landshoff
\cite{dl}. For larger values of $Q^2$ ($Q^2 > 1.0 GeV^2$) the usual
QCD based parametrisations of GRV or MRS \cite{grv, mrs} give a good
description of the data. 
\begin{figure}[htb]
\begin{center}
\mbox{\epsfig{file=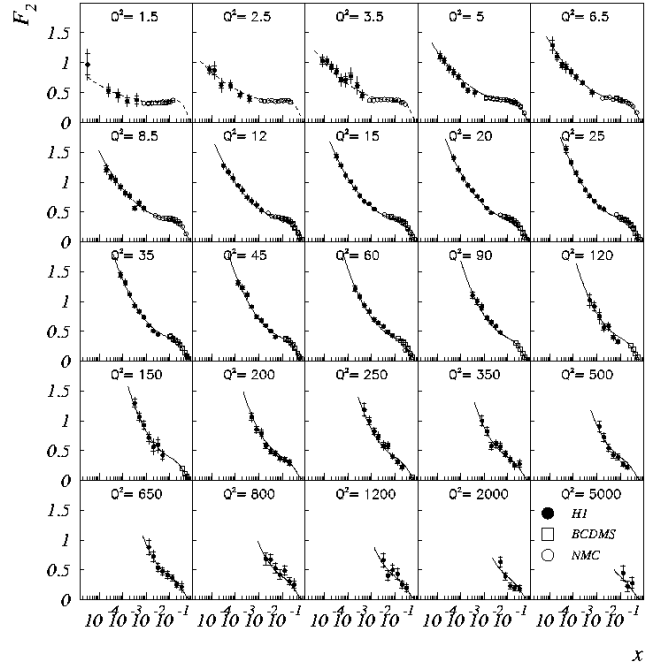,width=10truecm,angle=0}}
\caption{$F_2(x,Q^2)$ as a function of $x$. The curves are the NLO QCD
fit for $Q^2 \ge 5 GeV^2$. Curves below $5\ GeV^2$ are obtained by
backward evolution.(from \protect \cite{h1})}
\label{Fig 3.}
\end{center}
\end{figure}
\begin{figure}[htb]   
\begin{center}   
\mbox{\epsfig{file=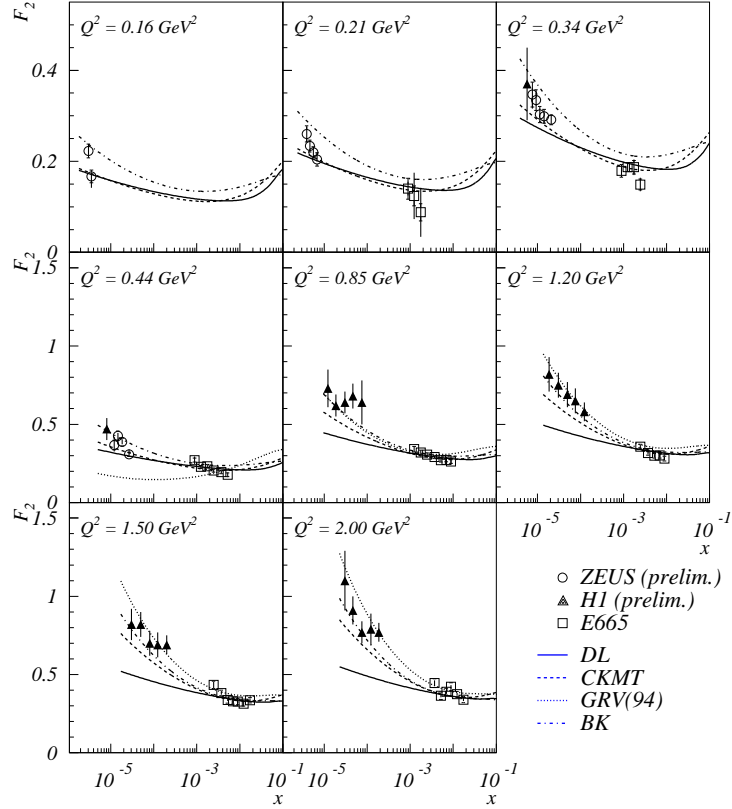,width=10truecm,angle=0}} 
\caption{$F_2(x,Q^2)$ as a function of $x$ for low $Q^2$
compared with model predictions.(from \protect \cite{halina})}
\label{Fig 4.}   
\end{center} 
\end{figure} 
\begin{figure}[htb]   
\begin{center}   
\mbox{\epsfig{file=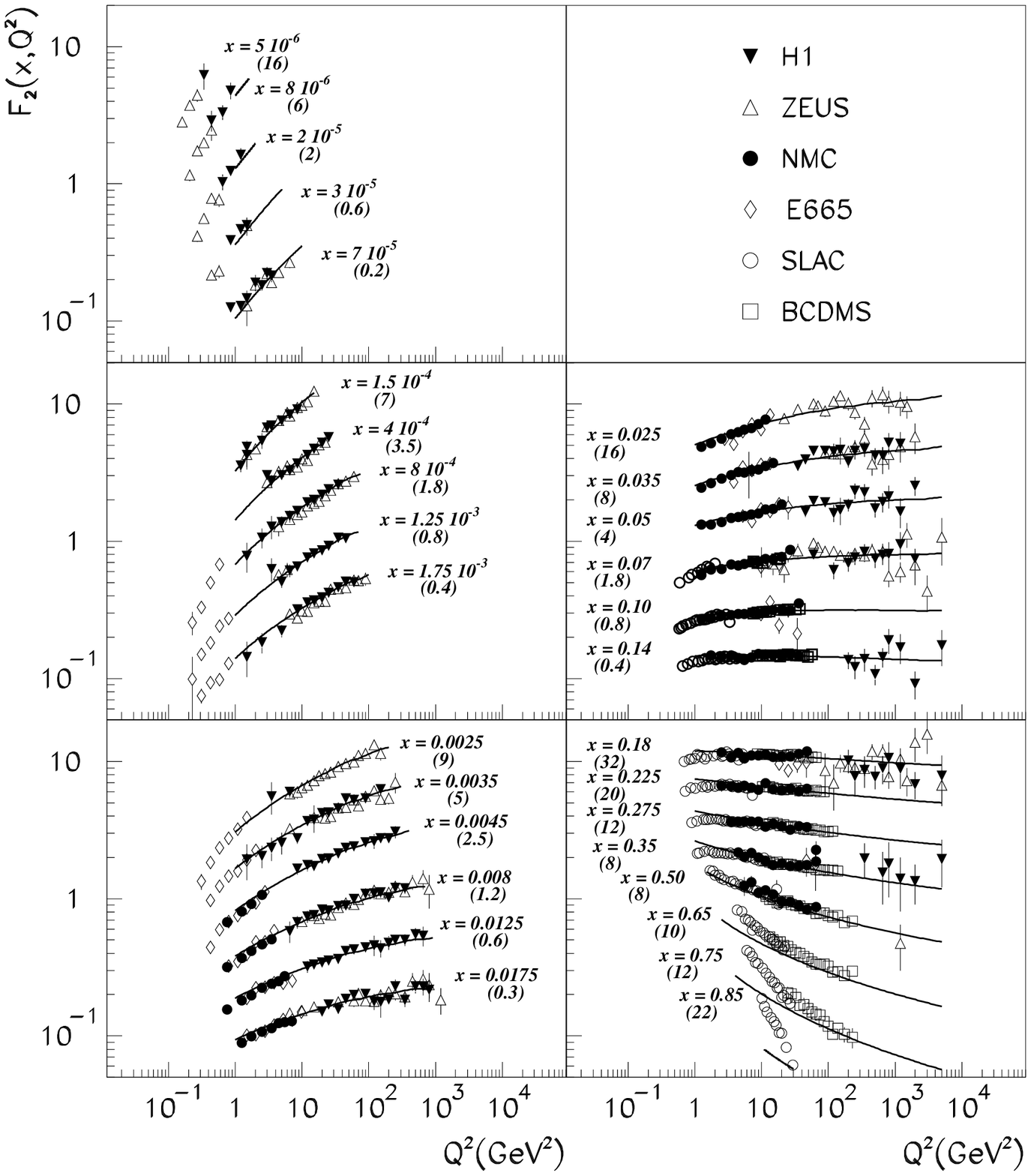,width=10truecm,angle=0}} 
\caption{Compilation of $F_2$ measurements as a function $Q^2$ for
selected values of $x$. The numbers in parenthesis are the scaling
factors by which the value of $F_2$ has been multiplied in the plot. The
curves are a NLO QCD fit by H1. (from \protect \cite{halina})}
\label{Fig 5.}   
\end{center} 
\end{figure} 

The behaviour of $F_2$ as a function of $Q^2$ for selected values of
$x$ is presented in figure 5. The growth of the structure function
with $Q^2$ for $x<0.01$ over three orders of magnitude in $Q^2$ is
clearly seen. With the new data from HERA the gap between fixed
target and the HERA experiments has been filled and the remarkable
agreement
with a NLO QCD fit performed by H1 over three decades in $Q^2$ and
over four decades in $x$ is clearly  seen. 

The H1 experiment has parametrised the rise of $F_2$ with $x$ by $F \sim
x^{-\lambda}$ at fixed $Q^2$ values. The result for $\lambda$ as a
function of $Q^2$ is shown in Figure 6. 
It is clear from the figure that $\lambda$ increases with increasing
$Q^2$ reaching a value of around $0.4-0.5$ around $Q^2 \sim 10^2-10^3
GeV^2$.
\begin{figure}[htb]   
\begin{center}   
\mbox{\epsfig{file=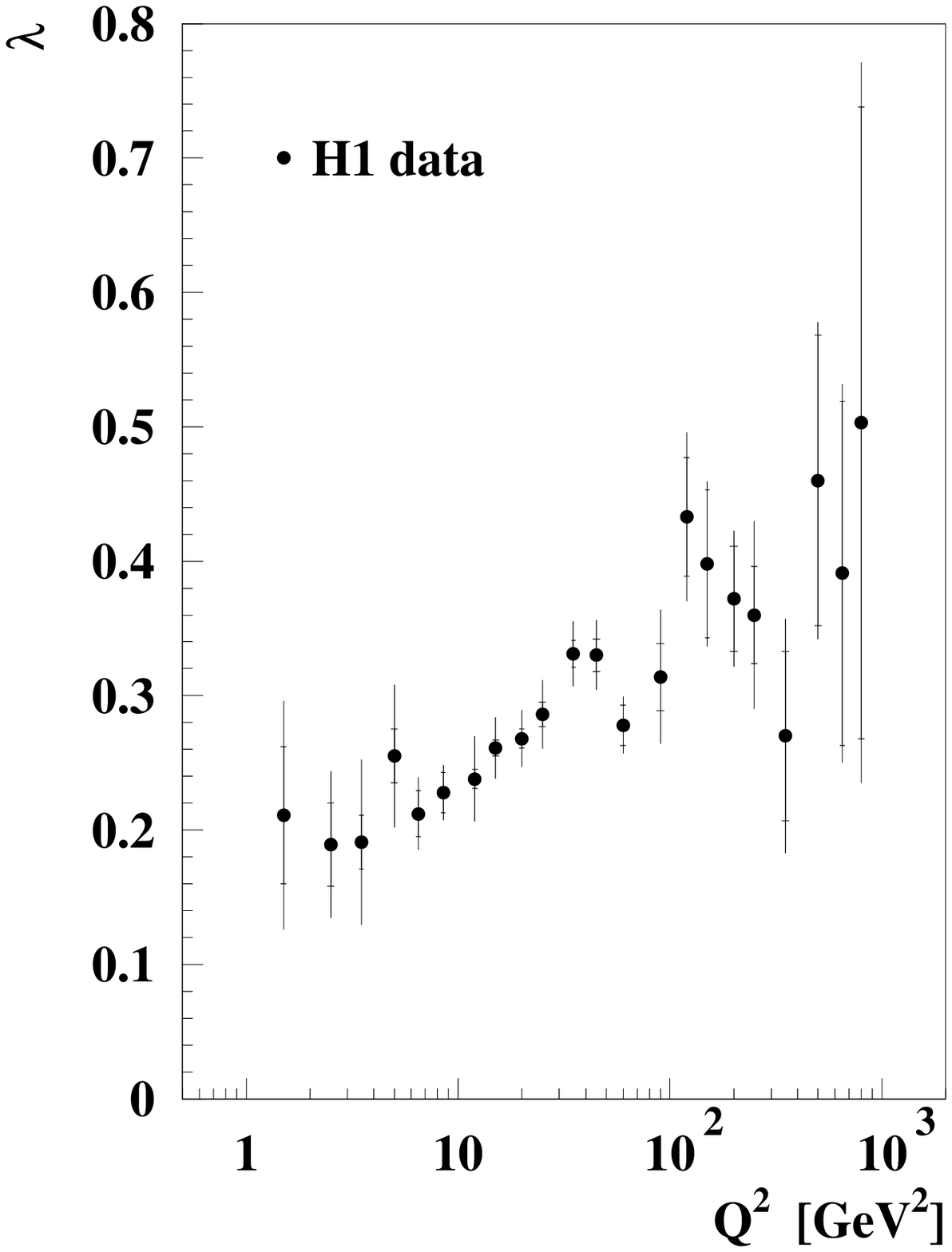,width=8truecm,angle=0}} 
\caption{Variation of the exponent $\lambda$ from fits of the form
$F_2 \sim x^{-\lambda}$ at fixed $Q^2$ values and $x < 0.1$.
(from \protect \cite{h1})}
\label{Fig 6.}   
\end{center} 
\end{figure} 

\section{The rise of $F_2$ and perturbative QCD}
The structure function $F_2(x,Q^2)$ is given in terms of the parton 
distributions
\begin{equation}
\frac{F_2(x,Q^2)}{x}\equiv\sum_{i=1}^{n_f}e_i^2C_i\otimes (q_i+\bar q_i)+C_g
\otimes g.
\end{equation}
The evolution of the structure function $F_2$ as a function of $Q^2$ is 
governed, in
perturbative QCD, by the Dokshitser-Gribov-Lipatov-Altarelli-Parisi
(DGLAP) equation \cite{dok, glap}, 
\begin{equation}
{\partial \over \partial \ln Q^2}
 \left(\begin{array}{c} q \\ g \end{array}\right)
 = {\alpha_s(Q^2) \over 2 \pi} \left[\begin{array}{cc}
 P_{qq} & P_{qg}  \\
 P_{gq} & P_{gg}
\end{array}\right] \otimes
\left(\begin{array}{c}q \\ g\end{array}\right) \, ,
\end{equation}
Here $q$ and $g$ are the quark and gluon distributions, $\otimes$ denotes 
convolution
with respect to $x$, i.e.$[f\otimes g](x)\equiv\int_x^1 {dy \over y}f({x\over
y})g(y)$, $n_f$ is the number of flavours of quarks, $e_i$ is the electric 
charge of
the quark $q_i$. The coefficient functions $C$'s at leading order are
\begin{equation}
C_i(x,Q^2)=\delta(1-x);\ \ \ C_g(x,Q^2)=0.
\end{equation}
At higher orders, they depend on the specific factorization scheme. In
the common parton scheme the above remains true to all orders. 

The splitting functions have the expansion
\begin{equation}
P_{ij}(x,Q^2)= \frac{\alpha_s}{2\pi} P_{ij}^{(1)}(x)+
\left(\frac{\alpha_s}{2\pi}\right)^2 P_{ij}^{(2)}(x)+ \ldots \, .
\end{equation}
The first two terms above define the NLO DGLAP evolution. 

The rise of $F_2$ as a function of $x$ is explained by the usual DGLAP 
evolution and
also by more non-conventional dynamics like the BFKL evolution \cite{bfkl}. 
Let us briefly review the two different approaches.

To understand the leading log summation technique in the DGLAP approach, 
we use the Dokshitzer method \cite{dok}. At low values of $x$, the gluon
is the dominant parton in the proton. He showed that the GLAP summation 
of the $(\alpha_s \log
Q^2)^n $ terms amounts to a sum of gluon ladder diagrams of the type shown 
in Figure
7 with $n$ gluon rungs. The form of the gluon splitting function 
$P_{gg}\sim 6/x$
for small $x$ gives the evolution of the gluon distribution function as
\begin{eqnarray}
xg(x,Q^2)&=&\sum_n(\frac{3\alpha_s}{\pi})^n\int^{Q^2}\frac{dk_{nT}^2}{k_{nT}^2}
\ldots
\int^{k_{3T}^2}\frac{dk_{2T}^2}{k_{2T}^2}\int^{k_{2T}^2}\frac{dk_{1T}^2}
{k_{1T}^2} \nonumber \\
&&\times \int_x^1\frac{d\xi_n}{\xi_n}\ldots\int_{\xi_3}^1\frac{\xi_2}{\xi_2}
\int_{\xi_2}^1\frac{\xi_1}{\xi_1}\xi_1g(\xi_1,Q_0^2)
\end{eqnarray}
\begin{figure}[htb]   
\begin{center}   
\mbox{\epsfig{file=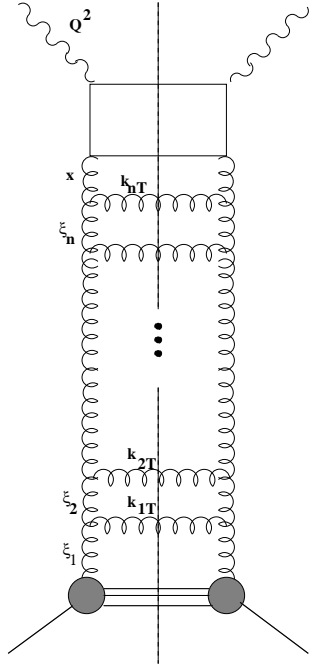,width=4cm,angle=0}} 
\caption{A gluon ladder diagram that contributes to the DLL or BFKL
ladder summations. }
\label{Fig 7.}   
\end{center} 
\end{figure} 
It is clear that the $(\log Q^2)^n$ builds up from the nested $k$ integrations. 
In fact this contribution comes from a region where the transverse momentum of
the emitted gluons are strongly ordered
\begin{equation}
Q^2 >> k_{nT}^2 >> \ldots >> k_{2T}^2 >> k_{1T}^2
\end{equation}
Similarly, the $(\log(1/x))^n$ term comes from the region where the longitudinal
momentum fractions are strongly ordered
\begin{equation}
x << \xi_n <<\ldots << \xi_2 << \xi_1.
\end{equation}
In the region of small $x$ the $\log(1/x)$ terms have also to be summed. 
With the assumption of $\xi_1g(\xi_1,Q_0^2)$ approaching a constant, 
say $G_0$ the result we get is
\begin{eqnarray}
xg(x,Q^2)& \sim &\sum_n\left(\frac{3\alpha_s}{\pi}\right)^n\left\{{1\over n!}
\left[\log\left(
{Q^2\over Q_0^2}\right)\right]^n\right\}\left\{{1\over n!}\left[\log\left(
{1\over x}\right)\right]^nG_0\right\} \\ \nonumber 
&& \sim G_0\exp\left\{2\left[\frac{3\alpha_s}{\pi}\log\left(Q^2\over Q_0^2\right)
\log \left(1\over x\right)\right]^{1\over 2}\right\}
\end{eqnarray}
in the limit of large $Q^2$ and small $x$.

This is an example of an all order double leading log (DLL) summation of 
$\alpha_s\log Q^2
\log(1/x)$ obtained by summing the strongly ordered gluon ladder diagrams 
as we have
just explained. It tells us that as $x$ decreases, $xg(x,Q^2)$ increases faster than 
any power of $\log(1/x)$ but slower than any power of $1/x$. 

If $k_{iT}\approx k_{i+1T}$ then we lose one power of $\log Q^2$. However, 
at small
$x$, this can be compensated by a large $\log(1/x)$ factor. Relaxing the strong
ordering constraint on the $k_T$'s and summing just the $\log(1/x)$ in the 
small $x$
region will give us the leading log summation in $\log(1/x)$ instead of DLL 
and we get
\begin{eqnarray}
xg(x,Q^2)& \approx& \sum_n(\frac{3\alpha_s}{\pi})^n\frac{1}{n!}\left[
c\ \log\left( {1\over x}\right)\right]^n \nonumber \\
&& \sim \exp\left[\lambda \log\left({1\over x}\right)\right] \nonumber \\
&&  \sim x^{-\lambda}
\end{eqnarray}
where $\lambda=(3\alpha_s/\pi)c$. This difficult summation was done 
rigorously by
Balitskij, Fadin, Kuraev and Lipatov (BFKL) \cite{bfkl}. The constant 
$c=4\ \log2$ and
so, for $\alpha_s \approx 0.2$, $\lambda\approx 0.5$. We have highly 
simplified the
discussion of the BFKL equation, and in practise one needs to work 
in terms of the
unintegrated gluon distribution $f(x,k_T^2)$ and integrate $f(x,k_T^2)/k_T^2$ 
to get the final gluon distribution.

We thus have two different predictions for the rise of $F_2$, the 
DLL (from the DGLAP
evolution equations) and the BFKL. In principle it should therefore be 
possible to
distinguish between these predictions by looking at the $F_2$ data from HERA. 
Unfortunately this is not that simple.  The steepness of the rise of $F_2$ with
decreasing $x$ can be controlled by varying $Q_0^2$, or the starting 
distribution 
$g(\xi_1,Q_0^2)$. In addition, the set of equations based only on 
$F_2$ measurements
is underconstrained and one needs one other measurement like the longitudinal
structure function $F_L$ in order to constrain the system fully. 

The net consequence of the above is that the present data on $F_2$ does not
distinguish betweem the BFKL and DGLAP (or for that matter the CCFM which 
embodies both) predictions. 

We would like to mention here a suggestion by Mueller
\cite{mueller} on measuring an observable that is less inclusive than the $F_2$
measurement (in which none of the properties of the hadronic final state is
measured). He suggests studying DIS events at small $x$ containing an identified
high-energy jet emitted close to the jet of proton fragments (see Figure 8). The
identified jet originates from the parton labelled $a$. If we study events with $x_j$
large ($\ge 0.05$) and $x$ very small ($\approx 10^{-4}$) the ratio $x/x_j$ will be
sufficiently small to reveal the $(x/x_j)^{-\lambda}$ behaviour of the 
BKFL ladder. 
Details may be found in the references mentioned above. This study is 
currently under way and preliminary results are reported in \cite{jet}.
\begin{figure}[htb]   
\begin{center}   
\mbox{\epsfig{file=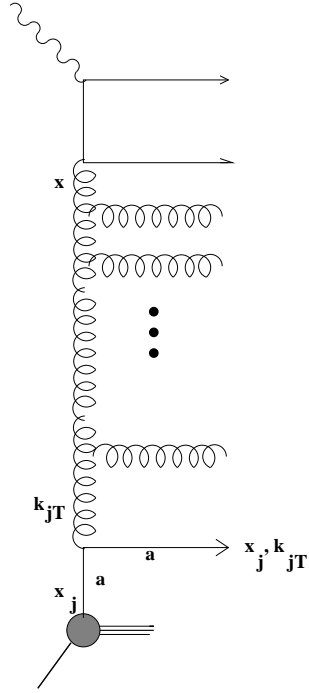,width=4truecm,angle=0}} 
\caption{A ladder diagram contributing to an identified forward jet as
discussed in the text}
\label{Fig 8.}   
\end{center} 
\end{figure} 

The behaviour of $F_2$ as a function of both $Q^2$ and $x$ (for large $Q^2$
and small $x$) can be neatly combined through the double
asymptotic scaling analysis of Ball and Forte \cite{bf}. They use the
two variables
\begin{equation}
\sigma\equiv \sqrt{\log(x_0/x)\log(\alpha_s(Q_0)/\alpha_s(Q))},\ \ 
\rho\equiv \sqrt{\frac{\log(x_0/x)}{\log(\alpha_s(Q_0)/\alpha_s(Q))}}
\end{equation}
with $\alpha_s(Q)$ the strong coupling constant evaluated upto two loops.

The parameters $x_0$ and $Q_0^2$ have to be determined experimentally.
$Q_0^2$ is found to be $2.5 GeV^2$  and $x_0 = 0.1$. (See, for example
\cite{h1}). $F_2$ is rescaled by the leading asymptotic factors
\begin{equation}
R_F(\sigma,\rho)=8.1 exp\left(-2\gamma\sigma+\omega{\sigma\over \rho} +
\frac{1}{2}\log(\gamma\sigma)+\log({\rho\over\gamma})\right)/\xi_F
\end{equation}
and
\begin{equation}
R_F^\prime(\sigma,\rho)=R_F exp(2\gamma\sigma)
\end{equation}
where
\begin{equation}
\xi_F=1+((\xi_1+\xi_2)\alpha_s(Q)-\xi_1\alpha_s(Q_0))(\rho/(2\pi\gamma)).
\end{equation}
The above constants are $\xi_1=(206n_f/27+6b_1/b_0)/b_0$, $\xi_2=13$,
$b_0=11-2n_f/3$, $\omega= (11+2n_f/27)/b_0$ and $b_1=102-38n_f/3$. Also,
$n_f$ is the number of flavours and $\gamma=\sqrt{12/b_0}$.

Thus the function $\log(R_F^\prime F_2)$ is predicted to rise linearly
with $\sigma$ and $R_F F_2$ is expected to be independent of $\rho$ and
$\sigma$.

Figure 9 shows the data for these two predictions. The first graph is
for $Q^2\ge 3.5 GeV^2$. Approximate scaling is observed for $Q\ge 5
GeV^2$ and $\rho\ge 2$. At high $\rho$ and low $Q^2$, the data tends to
violate the predicted scaling behaviour. The second graph shows the
appropriate data for $\rho\ge 2$ and $Q^2\ge 5 GeV^2$ as a function of
$\sigma$. The predicted linear rise is seen with slope of $2.50\pm
0.02\pm 0.06$ in good agreement with the QCD expectation of $2.4$ for
$4$ flavours.

\begin{figure}[htb]   
\begin{center}   
\mbox{\epsfig{file=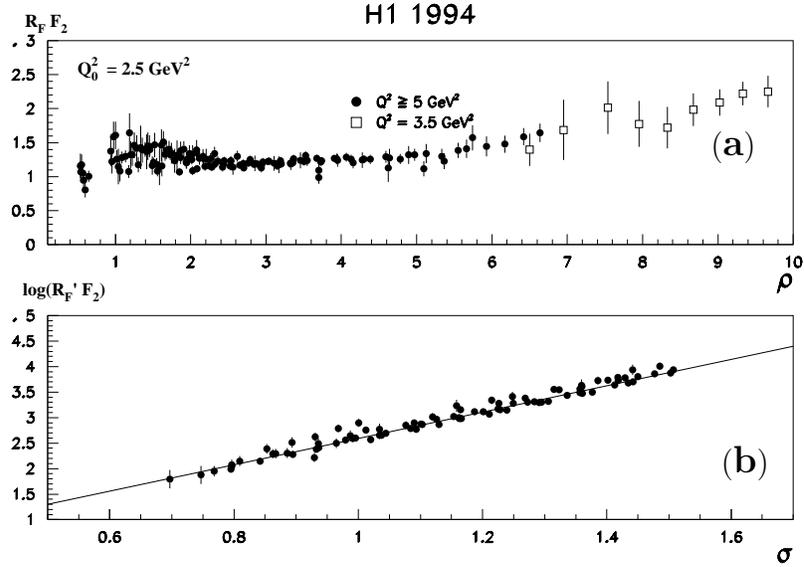,width=11truecm,angle=0}} 
\caption{The rescaled structure functions a) $R_FF_2$ vs. $\rho$ and b)
$\log(R_F^\prime F_2)$ vs. $\sigma$ as explained in the text. Only data
with $Q^2 \ge 5 GeV^2$ and $\rho > 2$ are shown in b).
(from \protect \cite{h1})}
\begin{picture}(0,0) \put(130,310){({\large \bf a})} \end{picture}
\begin{picture}(0,0) \put(130,190){({\large \bf b})} \end{picture}
\label{Fig 9.}   
\end{center} 
\end{figure} 

Thus in the low $x$ , high $Q^2$ region beyond $Q^2\ge 5 GeV^2$, scaling
is seen in the two variables $\rho$ and $\sigma$. The prediction of
double scaling is in very good agreement in this region with experiment.
Beyond the double scaling region, the evolution continues to be in
excellent agreement with the full NLO QCD evolution prediction.

One of the by-products of the measurement of the structure function
$F_2$ and its $\log Q^2$ evolution is that the gluon distribution inside
the proton has been measured with greater accuracy ({\it cf.} Figure 5
in \cite{halina}). 

To summarise this section, the structure function $F_2$ has been found
to rise steeply with decreasing $x$ down to $Q^2$ as low as $0.85
GeV^2$. This rise can be parametrised by a form $x^{-\lambda}$ where
$\lambda$ is found to be a function of $Q^2$ and reaches a value of
around $0.4-0.5$ around $1000 GeV^2$. This behaviour is described very
well by conventional NLO QCD (through the DGLAP equation). However this
does not rule out the existence of non-conventional dynamics like BFKL
because the system is underconstrained. Thus some other measurement
which is less inclusive than $F_2$ like the forward jet mentioned above,
or a measurement of the longitudinal structure function $F_L$ is
required. In any case, some new physics must take over at sufficiently
low $x$ otherwise the indefinite rise of $F_2$ would violate the
unitarity bound. A study of the longitudinal  structure function in the double
scaling limit may be found in \cite{sb1}.

\section{Large Rapidity Gap Events}
In approximately 10\% of the events used to study $F_2$, a large
rapidity gap in the hadronic final state was observed between the photon
and the proton fragmentation regions. This is described in Figure 10. 
These events were fairly independent of $W$ and $Q^2$. This implies
that there is no energy or colour flow between the struck parton and the
proton remnant direction. A generic explanation of these events is
through the exchange of a colour singlet parton which is typically a
pomeron. Such events are
typical of diffractive events through which one introduces the idea of
the pomeron. For our purposes, the pomeron is the generic name for a
particle the exchange of which is responsible for these large rapidity
gap events. 
\begin{figure}[htb]
\begin{center}
\mbox{\epsfig{file=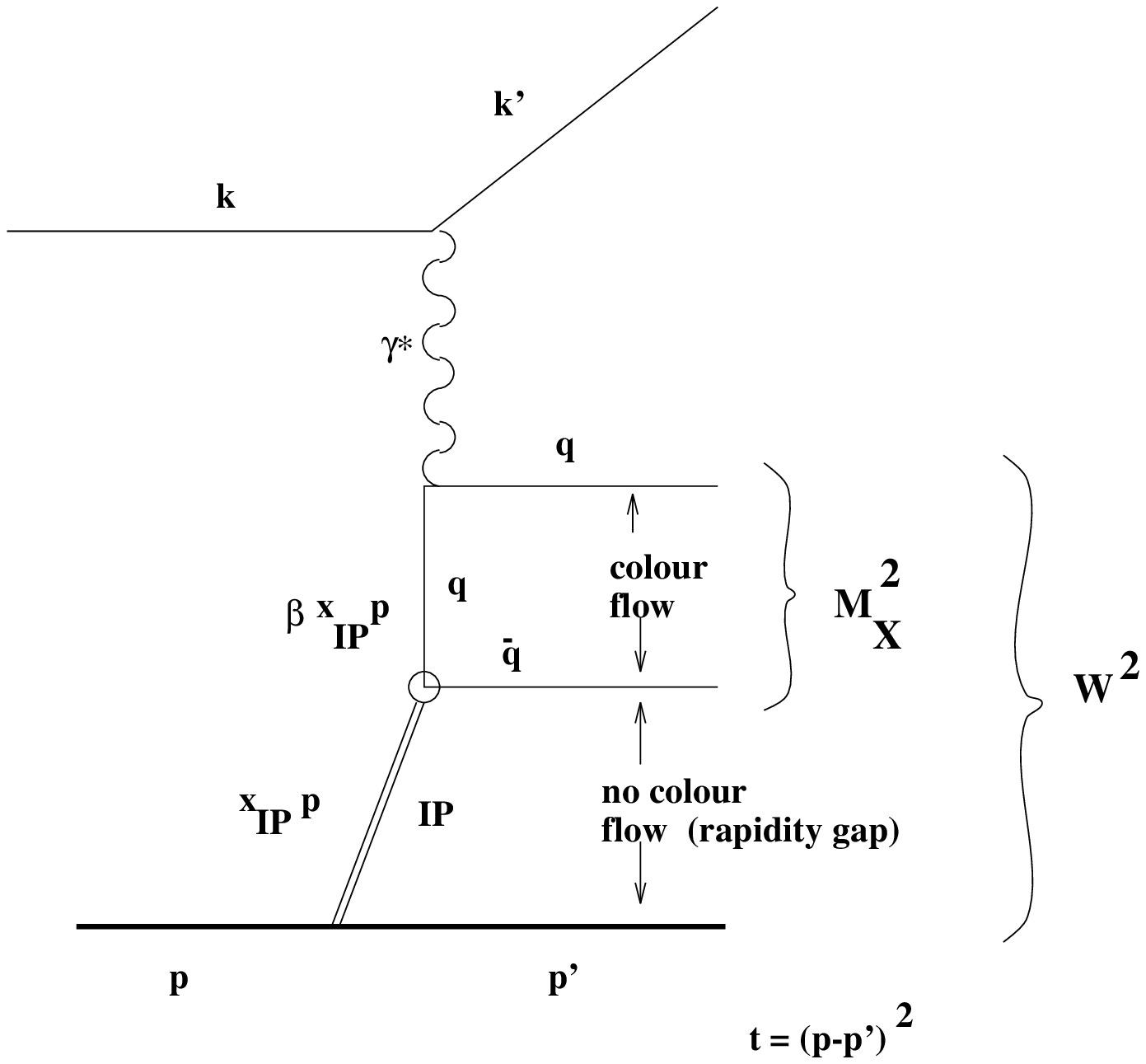,width=10truecm,angle=0}}
\caption{Schematic representation of a DIS event with a large rapidity
gap as discussed in the text}
\label{Fig 10.}
\end{center}
\end{figure}

One of the simplest ways to understand these processes is to consider
$\gamma^*p$ scattering in the rest frame of the proton. Here, the life
time of a $\bar q q$ fluctuation of the virtual photon is given by
$\tau\simeq 1/(2m_px)$ where $m_p$ is the mass of the proton. This time
scale corresponds to a distance of $c\tau\simeq 10^3 fm$ for
$x=10^{-4}$ which is much larger that the typical size of a hadronic
target. Thus the photon may fluctuate  into a $q\bar q$ pair before
arriving at the target, thereby making $\gamma^*p$ interaction similar
to hadron hadron interaction. Diffractive scattering emerges as a
consequence of the hadronic nature of the virtual photon.

It is important here to present the appropriate kinematic
variables for the process. We use the variables $x_\pom$ for the fraction
of the proton momentum carried by the generic pomeron $\pom$, and $\beta$
the fraction of the pomeron momentum carried by the struck quark with
$\beta=x/x_\pom$, $x$ being the usual Bjorken variable. Neglecting the
mass of the proton and taking $t\simeq 0$, (see Figure 10)
\begin{equation}
x_\pom = \frac{Q^2+M_X^2}{W^2+M_X^2};\ \ \ \ \beta=\frac{Q^2}{Q^2+M_X^2}
\end{equation}
Assuming the validity of Regge factorization for $\gamma^*p$ diffractive
interactions we have
\begin{equation}
\frac{d^2 \sigma(\gamma^*p)}{dt dx_{\pom}} = F_{\pom/p}(t,x_{\pom})
\sigma(\gamma^* \pom) \, ,
\end{equation}
where $ F_{\pom/p}(t,x_\pom)$ is the flux of the pomeron in the proton.
The DIS differential cross section for producing a diffractive hadronic
state $X$ of mass $M_X$ is
\begin{equation}
\frac{d^3\sigma^D}{dx_{\pom} d\beta dQ^2}=\frac{2\pi \alpha^2}{\beta
Q^4}
\left[ 1+(1-y)^2 \right]  F_2^{D(3)}(\beta,Q^2,x_{\pom}) \, ,
\end{equation}
$F_2^{D(3)}$ denotes the contribution of diffractive events with a given
$x_\pom$ integrated over $t$ to the proton structure function $F_2$ for
a given $Q^2$ and $x$. This part of the structure function is related to
the pomeron structure function $F_2^\pom$ by
\begin{equation}
F_2^{D(3)}(\beta,Q^2,x_{\pom}) = \frac{1}{x_{\pom}^n} F_2^{\pom}(\beta,Q^2) \, ,
\end{equation}
where $n=1-2 \alpha_{\pom}(0) + \delta_t$ independent of $\beta$ and
$Q^2$. 
\begin{figure}[ht]
\begin{center}
\mbox{\epsfig{file=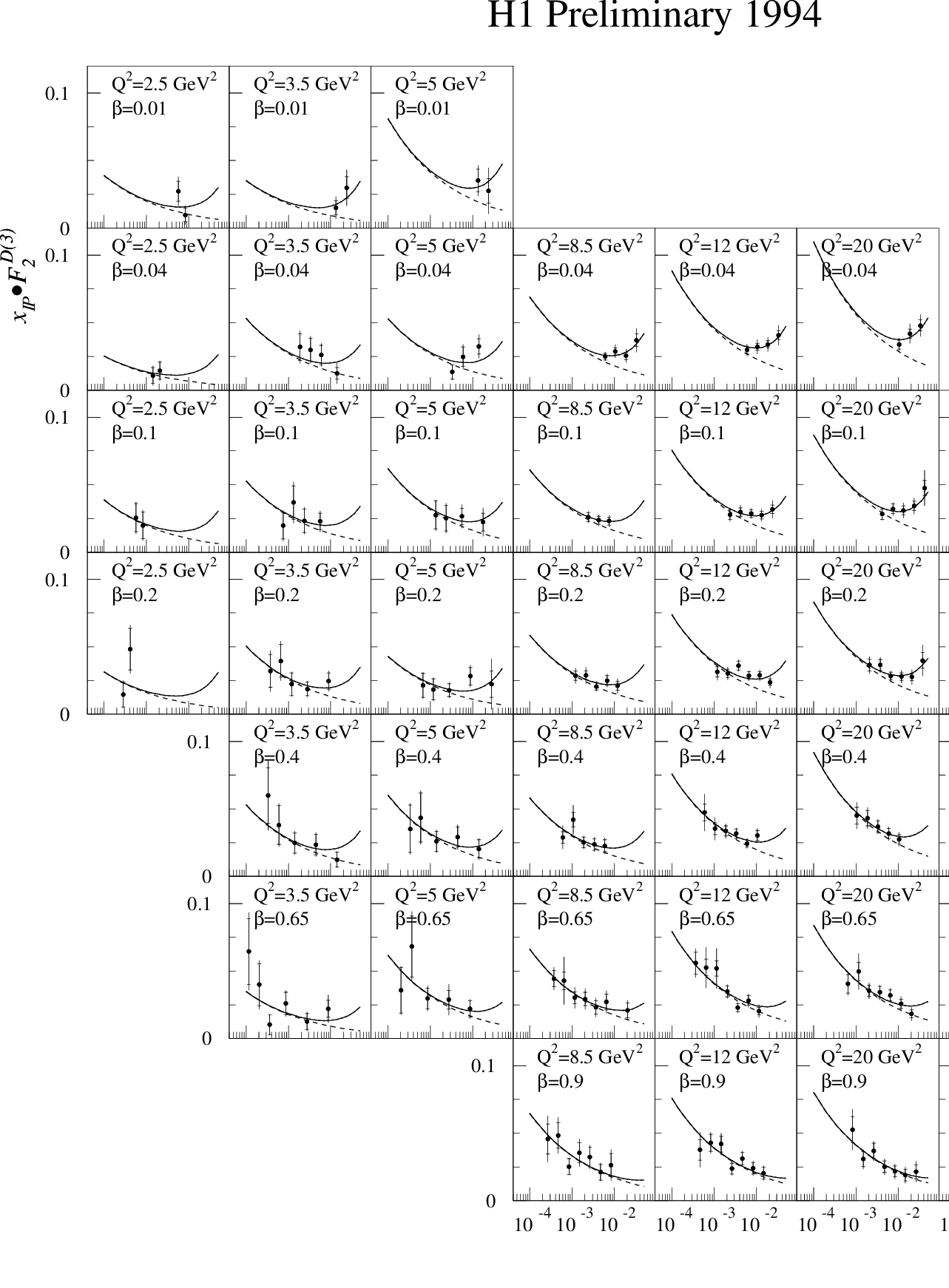,width=9truecm,angle=0}}
\caption{$x_\pom F_2^{D(3)}$ as a function of $x_\pom$ in bins of
$\beta$ and $Q^2$ as denoted in the figure. The full line is the fit of
an incoherent sum of the $\pom$ and $f$ meson trajectories. The dashed
line corresponds to the $\pom$ contribution. (from \protect
\cite{halina})}
\label{Fig 11.}
\end{center}
\end{figure}

H1 has carried out measurements of $x_\pom F_2^{D(3)}$ in the range
$2.5 < Q^2 < 65 GeV^2$, $0.01 < \beta < 0.9$ and $10^{-4} < x_\pom <
5.10^{-2}$ and this is shown in Figure 11. The pomeron trajectory fit
corresponds to an intercept
\begin{equation}
\alpha_\pom(0)=1.18\pm 0.02 \pm 0.07 
\end{equation}
which is somewhat higher than the intercept of a soft pomeron ($\sim
0.08$). The result from ZEUS is similar
\begin{equation}
\alpha_\pom(0)=1.17\pm 0.04\pm 0.08
\end{equation}
The quantity $\tilde{F}_2^D=\int F_2^{D(3)} dx_{\pom}$ is directly
proportional to $F_2^\pom$ and is found to be almost independent of
$Q^2$. This can be explained only by a large gluon contribution (see
\cite{halina} for details). In fact more than 80\% of the Pomeron
momentum is carried away by hard gluons in the range of $Q^2$ being
studied. 

To summarise, the pomeron, which is a generic term to describe a colour
singlet particle with vacuum quantum numbers, seems different in
different processes. In particular, the pomeron in these diffractive
events with large rapidity gaps seems to be harder than the usual soft
pomeron of Donnachie and Landshoff \cite{dl}. Using DGLAP evolution it
is found that the effective pomeron is made of mostly hard gluons.
It is clear however, that the pomeron is not a universal object and has
different intercepts in different processes.

\section{Photon Structure Function}
Even though a photon, being a gauge particle has no intrinsic structure,
its hadronic fluctuations (through its interaction with matter) give it a
partonic structure. Hard scattering processess (jets and charm production) yield
tests of perturbative QCD at NLO and give information about the quark and gluon
structure of the photon. We will very briefly touch upon some of these matters.
Large $E_T$ dijet photoproduction takes places at leading order through the 'direct'
scattering process ${\hat\sigma}(\gamma q\rightarrow qg)\otimes q^p$ and
${\hat\sigma}(\gamma q\rightarrow q{\bar q})\otimes g^p $ . There are also so-called
resolved processes (at higher order) ${\hat\sigma}(g q\rightarrow gq)\otimes 
g^\gamma\otimes q^p$ and ${\hat\sigma}(q q\rightarrow q q)\otimes q^p\otimes
q^\gamma$. The momentum fraction of the photon constituent can be
recontructed from the transverse energies and rapidities of the dijets,
\begin{equation}
x_\gamma= \frac{E_{T1}\exp(-\eta_1)+E_{T2}\exp(-\eta_2)}{2E_\gamma}
\end{equation}
Thus, 'direct' processes have $x_\gamma\approx 1$ and 'resolved' processes $x_\gamma
< 1$. The angular distribution of the dijets can also be measured to provide
information on the nature of the hard scattering process. In fact the expected
difference in distribution between direct and resolved process can be clearly seen in
the data and is again in good agreement with NLO QCD predictions. There are also a
host of other results both from H1 and ZEUS. A good summary of the results and their
theoretical implications can be found in the report of the photoproduction working
group of the HERA workshop on proton, photon and pomeron structure function
\cite{photon}.

\section{Exclusive vector meson production}
The detectors at H1 and ZEUS have the ability to study the nature of the
hadronic final state in detail. As a result there has been renewed
interest in the study of \jpsi production at HERA. The production cross
section for the process is dominated by photoproduction  i.e by the
interaction of almost real photons $Q^2\approx 0$. 

The process $\gamma p \rightarrow \jpsi X$ can take place through
various mechanisms. The elastic process corresponds to $X=p$ for which
there are various models proposed, in particular the soft pomeron
picture of Donnachie and Landshoff \cite{dl} and the Ryskin model
\cite{ryskin}.  The Ryskin model is based on perturbative QCD and is an
attempt to describe a hard pomeron through a gluon ladder of the BFKL
kind.  The elastic process can be with proton dissociation or without. 
In both cases, with small momentum transfer, the \jpsi particle retains
the full photon energy ($z\approx 1$ with $z=E_\psi/E_\gamma$ in the
proton rest frame). \jpsi production with proton dissociation also
leads to $z$ values close to one even though its strictly speaking, an
inelastic process. 
\begin{figure}[tb]
\begin{center}
\mbox{\epsfig{file=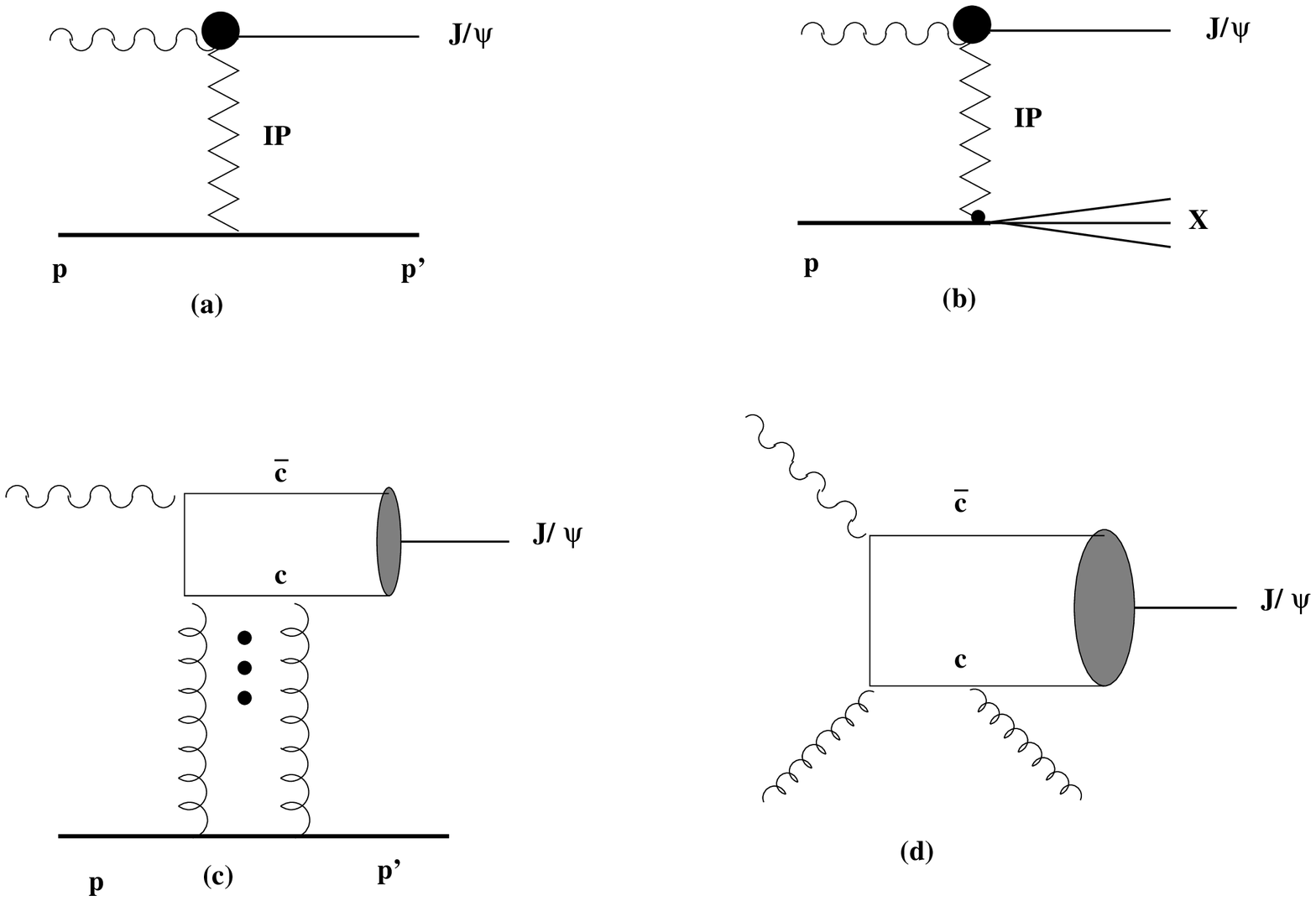,width=12truecm}}
\caption{\jpsi production mechanisms: a) elastic \jpsi production via
pomeron exchange b) diffractive proton dissociation c) elastic \jpsi
production in perturbative QCD (Ryskin model, see text) d) photon gluon
fusion model for inelastic \jpsi production in colour singlet model. }
\label{Fig 12.}
\end{center}
\end{figure}

However the principle inelastic process is that of photon-gluon fusion
which has been modelled by Berger and Jones \cite{bj} who calculated
this in perturbative QCD  due to the hard scale given by the mass of the
charm quark. In the colour singlet model for photoproduction of \jpsi
which is based on the Berger-Jones picture, the formation of the \jpsi
state is accompanied by the emission of a hard gluon. A comparison of
the colour singlet model to data gave a discrepancy in absolute
magnitude which was attributed to missing higher order calculations
(also called the "K-factor"). Subsequently, several improvements
including a complete NLO calculation \cite{kramer} have been performed
and compared successfully with the data. The possibility of colour
octet contributions have also been studied - for an overview, see, for
example, \cite{ck}. All these various processes are shown
diagramatically in Figure 12. 
There is not enough space to show the data for \jpsi production. 
The interested reader is referred to \cite{h1-jpsi} for data and experimental
details. We will only briefly describe the data.
The data show a rise with $W_{\gamma p}$ which in the HERA regime can
be modelled by $\sigma_{\gamma p}\propto W_{\gamma p}^\delta$. A fit to
the HERA data gives $\delta=0.64\pm 0.13$ (including systematic and
statistical errors). Using also the ZEUS data and lower energies, gives
$\delta=0.90\pm 0.06$.

The energy dependence from the Donnachie-Landshoff \cite{dl} soft
pomeron model corresponds to $\delta=0.32$. 
The Donnachie-Landshoff prediction falls well below
the data by a factor of 3 or so. The prediction of the QCD-based model
of Ryskin and including higher order corrections \cite{ryskin1} 
gives much better agreement in general.
The Ryskin model, based as it is on a 2 gluon
ladder, depends quadratically on the gluon distribution taken at
the scale $(Q^2+m_\psi^2)/4 \approx 2.4 GeV^2$. The distribution at this
scale can be parametrised by $xg(x) \propto x^{-\lambda}$. For values of
$x \approx m_\psi^2/W_{\gamma p}^2 \approx 10^{-3}$ the gluon
distribution from MRSA' \cite{mrsa} corresponds to $\lambda \approx 0.2$
which gives good agreement with data between 10 and 150 GeV. The GRV
parametrisation \cite{grv} corresponds to $\lambda\approx 0.3-0.4$
giving a curve with a somewhat steeper energy dependence than the data.

The inelastic data sample is selected by requiring $0.45\le z \le 0.90$ 
to reject elastic
(diffractive) and resolved photon events. The inputs are the charm mass
$m_c = 1.4 GeV$, $\Lambda_{{\bar MS}}=300 MeV$ and the renormalisation
scale is chosen to be the factorisation scale $\sqrt{2}m_c$. 
Once
again the gluon distributions  at low $x$ can be parametrised by
$xg(x) \propto x^{-\lambda}$ with $\lambda$ ranging from 0 (MRSD0')
\cite{mrsd0} to 0.4 (MRSG) \cite{mrsa}. Approximate agreement is seen,
however the data cover a wide range in absolute normalisation.  Since
the NLO corrections are not fully under control for $p_t^2\rightarrow 0$
and $z\rightarrow 1$, a restricted kinematic
sample with $z<0.8$ and $p_t^2 > 1 GeV^2$ has been studied by
experimentalists. There is improved agreement for
all the gluon distributions but the sensitivity is reduced.
Again, details may be found in \cite{h1-jpsi}. An application of the asymptotic form
of the gluon distribution as obtained from the double scaling limit to both the
inelastic and elastic regions may be found in \cite{sb2}.

To summarise, once again, the agreement with data is very good with
standard gluon parametrisations, provided a hard pomeron (as that built
up of gluon ladders as in the Ryskin model) is used. This picture 
explains the diffractive/elastic events far better than the soft
pomeron version of Donnachie and Landshoff \cite{dl}. In the elastic
case the Berger and Jones picture using the colour singlet model of
\jpsi production seems to give a satisfactory description of the data
once NLO corrections are included and suitable cuts on $z$ and $p_t^2$
are incorporated. 

\section{Conclusion}

The structure function $F_2$
has now been measured in the kinematic range $10^{-6}< x < 0.85$ and $0.1 < Q^2
< 30,000 GeV^2$. The rise of $F_2$ with decreasing  $x$ is well explained by
conventional DGLAP evolution but also by non-conventional dynamics like BFKL. A less
inclusive measurement is required to distinguish these different evolution equations.
The large rapidity gap events and diffractive scattering show that the (generic)
pomeron is differrent in different processes but is harder than the
Donnachie-Landshoff soft pomeron. Exclusive vector meson production also indicates
that both for inelastic and elastic/diffractive events, perturbative QCD based descriptions
explain the data fairly well. LEP and HERA have also given us the low $x$ regime of
the partonic structure of the photon, even though lack of space has not allowed me to
discuss these aspects.
The next frontier is that of polarised DIS at low $x$. The HERMES experiment is a 
step in that direction. 

The large body of theoretical work on low $x$ has given us a good understanding of
QCD dynamics at large parton densities. As a result, parton densities are now
available over a large range in $Q^2$ and $x$. The results from HERA have given us
new ways to connect partonic and hadronic degrees of freedom and contributed to a
major improvement in our understanding of the short distance structure of hadrons. 
Low $x$ QCD is one of the few areas of particle physics where  
theorists have had to work hard to keep up with experimentalists.

Due to constraints of space, I have left out many other aspects of HERA Physics, like
photon structure functions, charged current interactions, some aspects of jets in
photoproduction etc. The HERA Workshop on proton, photon and pomeron structure 
\cite{hera} discusses in detail many of the issues touched upon here as well as 
some of the other experimental results that have not been discussed here. \\
\vspace{2cm}

\noindent{\large \bf Acknowledgements}\\
I would like to thank the organisers of the XII DAE symposium on High Energy Physics
for inviting me to give this talk and for hospitality during the symposium.


\begin{thebibliography}{99}
\bibitem{h1}
H1 Collaboration; T. Ahmed et al., Nucl. Phys. {\bf B 429} (1994) 477;
S. Aid et al., Nucl. Phys. {\bf B 470} (1996) 3.
\bibitem{zeus}
ZEUS Collaboration; M. Derrick et al., Phys. Lett. {\bf B315} (1993) 481;
Z. Phys {\bf C69}, (1996) 607
\bibitem{dl}
A. Donnachie and P. V. Landshoff, Nucl. Phys {\bf B 231} (1983) 189;
 ibid. {\bf B244} (1984) 322; ibid. {\bf B267} (1986) 690; Phys. Lett.
 {\bf B296} (1992) 227; ibid {\bf B202} (1988) 131.
\bibitem{grv}
M. Gluck, E. Reya and A. Vogt, Z. Phys. {\bf C 67} (1995) 433.
\bibitem{mrs}
A. D. Martin, W. J. Stirling and R. G. Roberts, Phys. Rev. {\bf D 50}
(1994) 6734.
\bibitem{dok}
Y. L. Dokshitzer, Sov. Phys. JETP {\bf 46} (1977) 641 
\bibitem{glap}	
V. N. Gribov and L. N. Lipatov, Sov. J. Nucl. Phys. {\bf 15} (1972) 438, 675; 
G.  Altarelli and G. Parisi, Nucl. Phys. {B 126} (1977) 298.
\bibitem{bfkl}
V. Fadin, E. Kuraev, L. Lipatov, Sov. Phys. JETP {\bf 44} (1976) 443;
ibid. {\bf 45} (1977) 199; Y. Balitski and L. Lipatov, Sov. J. Nucl.
Phys. {\bf 28} (1978) 822.
\bibitem{mueller}
A. H. Mueller, Nucl. Phys. {\bf B} (Proc. Suppl.), {\bf 18C},(1990) 125 ;
J. Phys. {G 17}, (1991) 1443
\bibitem{jet}
Working Group Report on the structure of the proton, B. Badelek et al.,
J. Phys. {\bf G 22}, (1996) 815; in particular, see section 5.5 .
\bibitem{bf}
R. D. Ball and S. Forte, Phys. Lett. {\bf B 335} (1994) 77; Phys.
Lett. {\bf B 336} (1994) 77.
\bibitem{sb1}
T. Sarkar and Rahul Basu, hep-ph/9607232 - submitted to Phys. Lett. B
\bibitem{halina}
H. Abramowicz, Tests of QCD at Low $x$ - hep-ex/9612001 - to be
published in the proceedings of the 28th International Conference on
High Energy Physics, Warsaw, 1996.
\bibitem{photon}
Photoproduction Working Group Report, HERA Workshop on proton, photon and pomeron
structure function, J. Phys. {\bf G 22} (1996) 871.
\bibitem{ryskin}
M. G. Ryskin, Z. Phys. {\bf C 57} (1993) 89.
\bibitem{ryskin1}
M. G. Ryskin et al., preprint DTP/95/96, CBPF-NF-079/95, RAL-TR-95-065,
hep-ph/9511228 (1995), revised February 1996.
\bibitem{bj}
E.I. Berger and D.Jones, Phys. Rev. {\bf D 23}, (1981) 1521.
\bibitem{kramer}
M. Kramer, hep-ph/9508409, Nucl. Phys {\bf B 459} (1996) 3.
\bibitem{ck}
M. Cacciari and M. Kramer, hep-ph/9601276, DESY preprint DESY 96-005.
\bibitem{h1-jpsi}
H1 Collaboration; S. Aid et al., Nucl. Phys. {\bf B 472} (1996) 3.
(hep-ex/9603005)
\bibitem{mrsa}
A. D. Martin, R. G. Roberts and W. J. Stirling, Phys. Lett. {\bf B354}
(1995) 155.
\bibitem{mrsd0}
A. D. Martin, W. J. Stirling and R. G. Roberts, Phys. Lett. {\bf B306}
(1993) 145; ibid. {\bf B309} (1993) 492.
\bibitem{sb2}
T. Sarkar and Rahul Basu - hep-ph/9701337 - submitted to Phys. Lett. B.
\bibitem{hera}
HERA workshop on proton, photon and pomeron structure, J. Phys. {\bf G 22} (1996)
669.
\end{thebibliography}
\end{document}